\documentclass[letterpaper]{article}

\usepackage[preprint]{aaai2027}
\usepackage[hyphens]{url}
\usepackage{graphicx}
\usepackage{natbib}
\usepackage{caption}
\usepackage{booktabs}
\usepackage{multirow}
\usepackage{tabularx}
\usepackage{colortbl}
\usepackage{amsmath}
\usepackage{amssymb}
\usepackage{xspace}
\usepackage{enumitem}

\newcolumntype{Y}{>{\centering\arraybackslash}X}
\definecolor{smartrow}{RGB}{242,242,242}

\newcommand{\method}{SmartGR\xspace}

\title{SmartGR: Hierarchy and Beam-Aware Knowledge Distillation for Generative Recommendation}

\author{
    Ziheng Zhang\textsuperscript{\rm 1}\equalcontrib,
    Yu Cui\textsuperscript{\rm 1}\equalcontrib,
    Bohao Wang\textsuperscript{\rm 1},
    Yong He\textsuperscript{\rm 2},
    Chao Yu\textsuperscript{\rm 2},
    Chuan Yuan\textsuperscript{\rm 2},
    Wujie Sun\textsuperscript{\rm 1},
    Can Wang\textsuperscript{\rm 1},
    Jiawei Chen\textsuperscript{\rm 1}\corresponding
}
\affiliations{
    \textsuperscript{\rm 1}Zhejiang University, Hangzhou, China\\
    \textsuperscript{\rm 2}Ant Group, Hangzhou, China\\
    \{zhangziheng,cuiyu23,bohao.wang,sunwujie,wcan,sleepyhunt\}@zju.edu.cn\\
    \{heyong.h,tianjing.yc,yuanzheng.xy\}@antgroup.com
}
 
\begin{document}

\maketitle

\begin{abstract}
Generative recommendation (GR) has emerged as a promising paradigm for recommender systems.
Scaling up GR models can improve recommendation performance, but it also substantially increases inference cost.
Knowledge distillation provides a practical solution by transferring knowledge from a large GR model to a lightweight one.
However, existing distillation methods do not account for two GR-specific challenges: imbalanced distillation difficulty across the semantic ID (SID) hierarchy and incorrect prefix pruning during beam search.
To address these challenges, we propose \method, a novel distillation framework that utilizes Hierarchy-Aware SID Distillation to transfer the teacher's modeling capability across the hierarchy and leverages Beam-Aware Ranking Distillation to distill the teacher's ranking preferences during beam search.
Extensive experiments on four benchmark datasets demonstrate the effectiveness and efficiency of \method, improving the performance by 8.6\% while achieving a $2.39\times$ inference speedup on average.
 \end{abstract}

\section{Introduction}
\label{sec:introduction}

Generative recommendation (GR) has recently attracted growing attention as an emerging paradigm for recommender systems~\cite{geng2022p5,rajput2023tiger,wang2024letter,liu2024etegrec}.
Unlike traditional sequential recommenders~\cite{hidasi2016gru4rec,kang2018sasrec,sun2019bert4rec}, GR represents each item with a semantic ID (SID), providing a natural way to incorporate item semantics and scale to large item space~\cite{singh2024semanticids,rajput2023tiger,ju2025grid}.
A typical GR pipeline comprises two stages: (i) item tokenization, which maps each item to a hierarchical SID; and (ii) autoregressive generation, which generates the target SID from the user history and uses beam search to rank candidate SID paths.

Generative recommendation can well enjoy the scaling law, with larger models generally achieving better recommendation performance~\cite{zhai2024hstu,zhou2025openonerec,kong2025minionerec}.
However, larger models require more computation and lead to higher inference latency~\cite{lin2025atspeed,guo2026sidmlp}.
Table~\ref{tab:onerec-scale-tradeoff} compares the recommendation performance and total inference time of the typical GR model OneRec~\cite{zhou2025openonerec} with 1.7B and 8B parameters on Kuaishou Ad and Video.
Increasing the model scale indeed improves recommendation performance, but this improvement comes at the cost of significantly increasing inference time by approximately 60\%--90\%, which is unacceptable for real-world recommendation systems.
Consequently, inference acceleration is essential for practical generative recommendation~\cite{lin2025atspeed,guo2026sidmlp}.

\begin{table}[t]
\centering
\normalsize
\setlength{\tabcolsep}{1.5pt}
\begin{tabular*}{\columnwidth}{@{\extracolsep{\fill}}llccc@{}}
\toprule
\textbf{Dataset} & \textbf{Model} & \textbf{Pass@32} & \textbf{Recall@32} & \textbf{Time (s)} \\
\midrule
Ad & OneRec-1.7B & 21.24 & 7.39 & 694.87 \\
& OneRec-8B & 27.19 & 9.75 & 1132.03 \\
\cmidrule{2-5}
& Gain & $+28.0\%$ & $+31.9\%$ & $1.63\times$ \\
\midrule
Video & OneRec-1.7B & 16.95 & 2.74 & 2157.33 \\
& OneRec-8B & 20.90 & 3.68 & 4037.11 \\
\cmidrule{2-5}
& Gain & $+23.3\%$ & $+34.3\%$ & $1.87\times$ \\
\bottomrule
\end{tabular*}
\caption{Performance and inference latency comparison on Kuaishou Ad and Video datasets under the representative GR model OneRec with the parameter scale of 8B and 1.7B.}
\label{tab:onerec-scale-tradeoff}
\end{table}
 
Knowledge distillation (KD)~\cite{hinton2015distilling,gu2024minillm} is a promising model compression technique that transfers knowledge from a large model (teacher) to a compact model (student), and has been widely applied to traditional recommender models to reduce inference latency~\cite{tang2018rd,lee2019cd}.
Existing recommendation distillation methods mainly transfer teacher item rankings~\cite{tang2018rd,lee2019cd,reddi2021rankdistil,zhu2026rcekd}, intermediate representations~\cite{romero2015fitnets,cui2024dllm2rec,zhu2025pckd}, or collaborative signals~\cite{kang2021topology,forouzandeh2025sharp} into a compact student model.
These methods are largely designed to enhance traditional sequential recommendation models with auxiliary knowledge and capabilities from large teacher models.

Although KD has proven effective for traditional recommendation, GR differs in its hierarchical SID-based item representation and beam-search inference mechanism, which introduce the following two challenges for distillation:

\textbf{Challenge 1: imbalanced distillation difficulty across the SID hierarchy.}
GR follows a hierarchical generation process in which each SID token is predicted conditionally on the previously generated prefix.
Meanwhile, SIDs follow a coarse-to-fine semantic hierarchy~\cite{rajput2023tiger,wang2024letter,ju2025grid}, which means that the distillation difficulty is imbalanced across SID positions.
Distillation must therefore capture the fine-grained conditional distributions along the SID sequence.
Table~\ref{tab:sid-prefix-recall} reports the teacher's recommendation performance gains over the student across the SID hierarchy.
It shows that the teacher's advantage becomes more pronounced at deeper SID levels.
Consequently, GR distillation needs to be hierarchy-aware rather than treating all SID levels uniformly.

\begin{table}[t]
\centering
\normalsize
\setlength{\tabcolsep}{2pt}
\begin{tabular*}{\columnwidth}{@{\extracolsep{\fill}}lccc@{}}
\toprule
\textbf{Dataset} & \textbf{L1} & \textbf{L1$\to$L2} & \textbf{L2$\to$L3} \\
\midrule
Amazon Beauty & $-0.42$ & $+2.73$ & $+4.12$ \\
Amazon Toys   & $+0.83$ & $+1.79$ & $+4.07$ \\
Kuaishou Video & $+0.75$ & $+4.82$ & $+1.79$ \\
Kuaishou Ad    & $+0.77$ & $+3.82$ & $+6.35$ \\
\bottomrule
\end{tabular*}
\caption{Performance gains (\%) of OneRec-8B over the OneRec-1.7B across the SID hierarchy. OneRec-8B achieves higher conditional prefix retention at deeper SID transitions.}
\label{tab:sid-prefix-recall}
\end{table}
 
\begin{figure}[!t]
  \centering
  \captionsetup{font=small,labelfont=normalfont,textfont=normalfont}
  \includegraphics[width=\columnwidth]{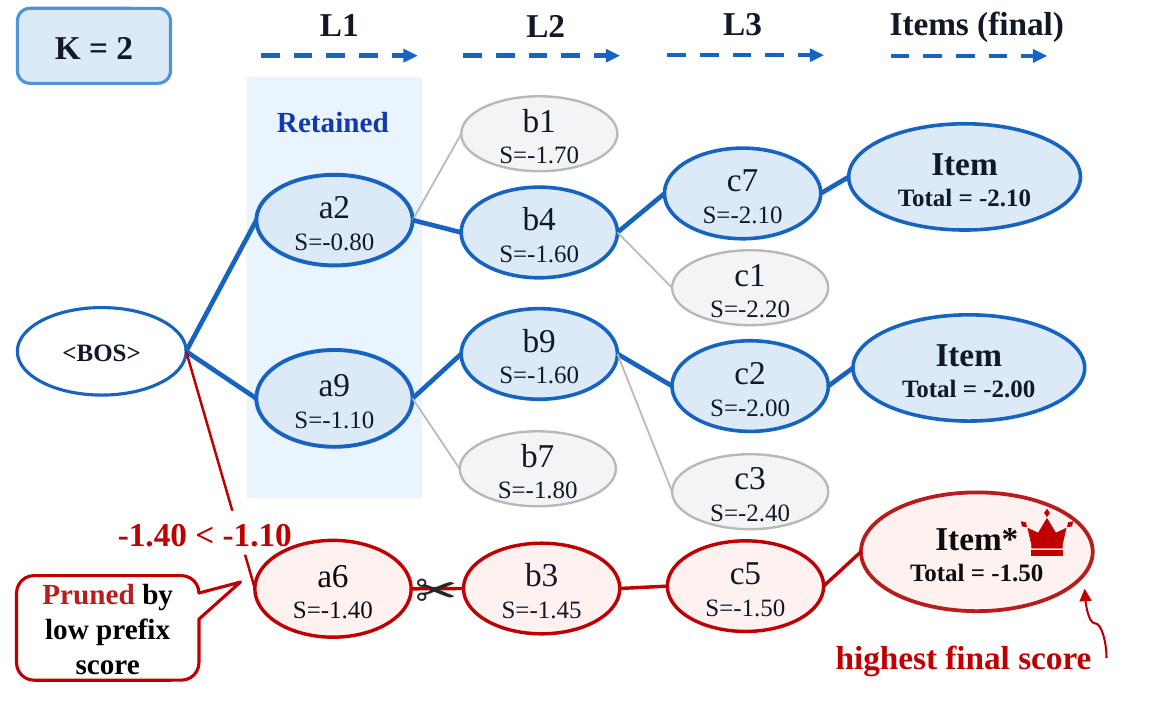}
{\small (a)}
\includegraphics[width=\columnwidth]{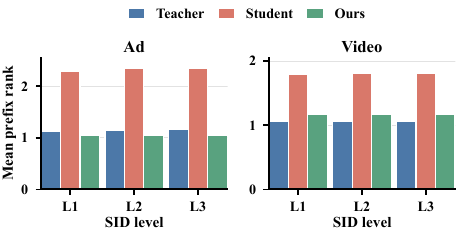}
{\small (b)}
\caption{(a) Beam-search pruning with beam width \(K=2\), where retained paths are shown in blue and the pruned target path is shown in red; the item with the highest final score can be removed when an intermediate prefix falls below the beam threshold. (b) Mean prefix rank of each model's global Top-1 item at each SID level; the teacher ranks the item better than the student, while Ours consistently improves the student.}
  \label{fig:positive-beam-pruning}
\end{figure}

\textbf{Challenge 2: incorrect prefix pruning during beam search generation.}
During inference, beam search generates multiple candidate item SIDs and retains only the Top-$K$ prefixes according to their cumulative scores at each decoding step.
Consequently, an item that would receive a high final score under the trained model may still be pruned during beam-search inference because an intermediate prefix has a low score~\cite{yu2026apao,yang2026bear}.
This stepwise beam-search inference creates a distillation challenge: aligning teacher distributions at individual SID positions does not guarantee that a teacher-preferred item path remains among the student's Top-$K$ prefixes, as shown in Figure~\ref{fig:positive-beam-pruning}(a).
We therefore compare the prefix-ranking capabilities of the teacher and student.
Figure~\ref{fig:positive-beam-pruning}(b) shows that the teacher achieves lower mean prefix ranks for the global Top-1 item, indicating stronger ranking capability during beam search.
This result shows that GR distillation needs to account for the teacher's ranking preferences during beam search.

\begin{figure*}[!t]
  \centering
  \includegraphics[width=0.97\textwidth]{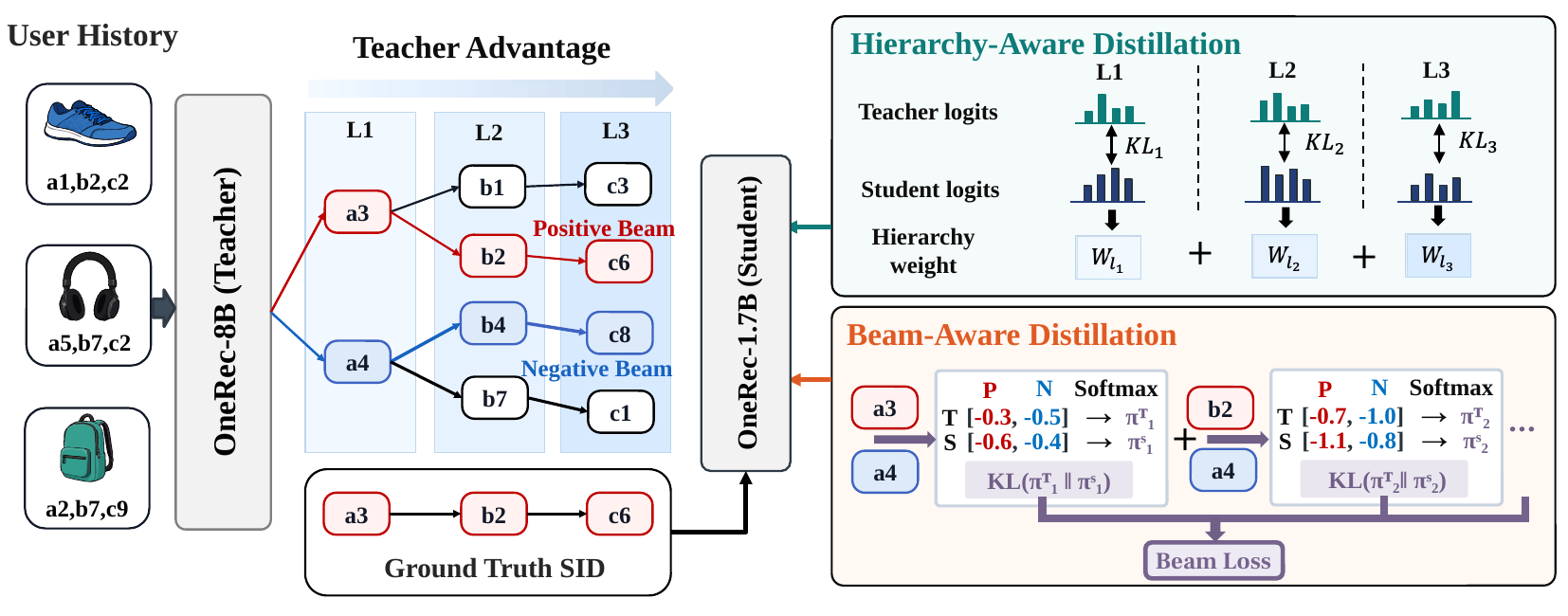}
\caption{Overview of the \method framework.}
  \label{fig:grdistill-overview}
\end{figure*}

To address these challenges, we propose \textbf{\method}, a novel \textbf{\underline{S}}ID hierarchy and bea\textbf{\underline{m}}-se\textbf{\underline{a}}rch \textbf{\underline{r}}anking dis\textbf{\underline{t}}illation framework tailored to \textbf{\underline{G}}enerative \textbf{\underline{R}}ecommendation.
\method transfers teacher knowledge through two distillation strategies:
(i) \textbf{Hierarchy-Aware SID Distillation} assigns learnable hierarchy-aware weights to SID levels, allowing the distillation strength to adapt across SID positions and addressing the imbalanced distillation difficulty across the SID hierarchy.
(ii) \textbf{Beam-Aware Ranking Distillation} uses the teacher's ranking preferences during beam search by modeling the relative order between cumulative prefix scores, reducing the risk that an item with a high final score is pruned because of a low-scoring prefix.
Together, these two objectives effectively transfer the hierarchical SID modeling and beam-search ranking capabilities from a larger teacher GR model to a smaller student.
Experiments across recommendation datasets from multiple domains demonstrate the effectiveness of \method, improving recommendation performance by 8.6\% while achieving a $2.39\times$ inference speedup on average.

In summary, our contributions are as follows:
\begin{itemize}
  \item We highlight the importance of knowledge distillation for GR and reveal its two core challenges: imbalanced distillation difficulty across the SID hierarchy and incorrect prefix pruning during beam search.
  \item We propose \method, which utilizes Hierarchy-Aware SID Distillation to transfer the teacher's modeling capability across the SID hierarchy, and leverages Beam-Aware Ranking Distillation to distill the teacher's ranking preferences during beam search generation.
  \item Extensive experiments on four benchmark datasets demonstrate that \method significantly improves recommendation accuracy while being highly cost-effective.
\end{itemize}
 \section{Preliminary}
\label{sec:preliminary}

\subsection{Generative Recommendation}
This paper focuses on generative recommendation (GR), which formulates next-item recommendation as an autoregressive generation problem in a tokenized item space, rather than directly scoring candidate items~\cite{rajput2023tiger,wang2024letter,ju2025grid}.
Let $\mathcal{I}$ denote the item catalog.
For a user $u$, let $H_u=(i_1,i_2,\ldots,i_n)$ denote the chronologically ordered interaction history, where each $i_j$ is an item from $\mathcal{I}$, and let $i_{n+1}$ be the next item to predict.
A standard GR model typically consists of the following steps:

\paragraph{1) Item tokenization.}
Item tokenization encodes the item semantic information into a fixed-length sequence of structured semantic ID (SID), which can be denoted by $\phi(i)=(z_{i,1},\ldots,z_{i,L})$, where $L$ is the SID length and $z_{i,\ell}$ is the SID token at position $\ell$.
For the tokenizer $\phi(\cdot)$, RQ-VAE~\cite{lee2022rqvae} and RQ-Kmeans~\cite{zhou2025openonerec} are typically used to construct SIDs through residual quantization, where later levels quantize the residual left by earlier levels, \emph{i.e., }  earlier SID tokens represent coarse item groups, while later tokens progressively distinguish finer-grained groups.

\paragraph{2) Autoregressive Generation.}
After tokenization, a user’s historical behavior can be represented as the item SID sequence $X_u=[\phi(i_1);\ldots;\phi(i_n)]$, which is then constructed as the model input $x$.
For the target item SID $\mathbf{g}=\phi(i_{n+1})=(g_1,\ldots,g_L)$, the model factorizes its probability as:
\begin{equation}
p_{\theta}(\mathbf{g}\mid x)
=
\prod_{\ell=1}^{L}
p_{\theta}(g_\ell\mid x,\mathbf{g}_{<\ell}),
\label{eq:gr-autoregressive}
\end{equation}
where $\theta$ is the model parameters and $\mathbf{g}_{<\ell}=(g_1,\ldots,g_{\ell-1})$ denotes the target SID prefix before level $\ell$.
The target item is recovered only when the complete SID is generated correctly.

\paragraph{3) Beam Search Decoding.}
During inference, autoregressive generation commonly uses beam search to produce a ranked set of candidate items~\cite{rajput2023tiger,xu2026trierec,yu2026apao,yang2026bear}.
Starting from the root of the SID prefix tree, the model expands each retained prefix with  next-level SID tokens and scores a prefix $\mathbf{z}_{1:\ell}=(z_1,\ldots,z_\ell)$ by its cumulative log-probability,
\begin{equation}
S_{\theta}(\mathbf{z}_{1:\ell}\mid x)
=
\sum_{j=1}^{\ell}
\log p_{\theta}(z_j\mid x,\mathbf{z}_{<j}).
\label{eq:gr-prefix-score}
\end{equation}
At each SID level, only the $K_{\mathrm{beam}}$ valid prefixes with the highest cumulative scores are retained.

\subsection{Knowledge Distillation for Recommendation}

Knowledge distillation (KD) transfers knowledge from a large teacher to a compact student for model compression~\cite{hinton2015distilling,gu2024minillm}.
Traditional KD for recommendation  methods  mainly transfer the knowledge like item rankings~\cite{tang2018rd,lee2019cd}, intermediate representations~\cite{zhu2025pckd,zhu2025freqd}, or collaborative structures~\cite{kang2021topology}.
Recent studies extend KD to LLM-based recommendation by transferring complete-item outputs or representations~\cite{cui2024dllm2rec,chen2025c2kd}.
Although effective, these KD strategies cannot be directly transferred to GR due to fundamental differences in model mechanisms. GR represents each item as a short-length and coarse-to-fine SID and retrieves items through beam search decoding. However, traditional recommenders represent each item with a single ID and directly predict the next item based on logits, whereas LLM-based recommenders process long, variable-length natural-language token sequences instead of fixed-format SID.

Recent studies explore KD for GR along two directions.
One direction modifies the student architecture to reduce autoregressive decoding cost, but increases the complexity of model design and training~\cite{guo2026sidmlp}.
The other transfers codeword distributions and complete candidate ordering to improve recommendation quality, but does not explicitly identify where the large teacher outperforms the lightweight model~\cite{xie2025lohrec}.
In contrast, \method identifies the large teacher's advantages in hierarchical SID modeling and cumulative prefix ranking during beam search, and transfers them using only two simple loss functions.

 \section{Methodology}
\label{sec:method}

In this section, we introduce \method, a distillation framework for GR that addresses imbalanced distillation difficulty across the SID hierarchy and incorrect prefix pruning during beam search. 
Given a training context and its target SID, \method uses Hierarchy-Aware SID Distillation to assign a learnable hierarchy-aware weight to the distillation loss at each SID level.
Beam-Aware Ranking Distillation further transfers the teacher's ranking preference between a positive and a negative beam based on their cumulative prefix scores.
The overall framework is illustrated in Figure~\ref{fig:grdistill-overview}.

\subsection{Hierarchy-Aware SID Distillation}
\label{sec:hierarchy-aware-sid-distillation}

SID levels follow a coarse-to-fine hierarchy and have different distillation difficulties, so assigning the same distillation strength to every position cannot account for how the teacher's advantage varies across the hierarchy.
As shown in Table~\ref{tab:sid-prefix-recall}, the teacher's advantage becomes more pronounced as the SID hierarchy deepens and varies across datasets.
We therefore introduce a hierarchy-aware distillation objective, named SID loss, which uses SID depth as a structural prior to learn distillation weights across SID positions from data.

Given the cached teacher beams $\mathcal{B}(x)=\{b_k\}_{k=1}^{K_{\mathrm{beam}}}$ sorted in descending order by teacher score, let $b_k=(z_{k,1},\ldots,z_{k,L})$ and $s_k$ denote the $k$-th beam and its score, respectively, with $s_1\geq\cdots\geq s_{K_{\mathrm{beam}}}$.
We write $\mathbf{z}_{k,<\ell}=(z_{k,1},\ldots,z_{k,\ell-1})$ for its SID prefix before level $\ell$.
We select the teacher beam $b_{k^*}$ that shares the longest SID prefix with the target SID; when multiple beams have the same shared-prefix length, we choose the highest one.
Let $L^*$ denote this longest common-prefix length and $\mathcal{L}(x)=\{1,\ldots,L^*\}$ denote the corresponding SID levels.
We restrict distillation to $\mathcal{L}(x)$ because only these overlapping positions provide teacher information consistent with the target SID.
When $L^*=0$, we skip distillation for this sample.

For the $\ell$-th valid SID level of the selected teacher beam $b_{k^*}$, the student distribution vector under the same prefix is
\begin{equation}
\mathbf{p}_S^\ell
=
p_S\!\left(\cdot\mid x,\mathbf{z}_{k^*,<\ell}\right).
\end{equation}
The corresponding teacher distribution vector is denoted by $\mathbf{q}_T^\ell$, and SID loss matches it with $\mathbf{p}_S^\ell$.
To determine the distillation strength at this SID level, we map the normalized SID depth $d_\ell=\ell/L$ to a learnable level score:
\begin{equation}
a_\ell=f_{\theta}(d_\ell),
\qquad
w_\ell(x)
=
\frac{\exp(a_\ell)}
{\sum_{j=1}^{L^*}\exp(a_j)},
\end{equation}
where $w_\ell(x)$ is the normalized distillation weight at SID level $\ell$ and $f_{\theta}$ is a learnable function shared across training samples.
Concretely, we instantiate $f_{\theta}$ as a normalized monotone $\tanh$-based function:
\begin{equation}
f_{\theta}(d)
=
\frac{\tanh\!\left(\theta d / \tau_{\mathrm{lev}}\right)}
{\tanh\!\left(\theta / \tau_{\mathrm{lev}}\right)},
\qquad
\theta
=
\theta_{\max}\sigma\!\left(\theta_{\mathrm{raw}}\right),
\end{equation}
where $\theta_{\mathrm{raw}}$ is a scalar learned jointly with the student, $\theta_{\max}>0$ bounds $\theta$, $\tau_{\mathrm{lev}}>0$ is a fixed scaling hyperparameter, and $\sigma(\cdot)$ is the logistic function.
Results with other weighting functions are reported in Appendix~\ref{app:sid-weighting-ablation}.
This function maps normalized SID depths to $[0,1]$ and increases monotonically, so deeper SID levels receive no smaller weights.
The learned $\theta$ controls the shape of the depth-to-weight mapping.
SID loss applies the hierarchy-aware weights to the KL divergence between the teacher and student conditional SID distributions:
\begin{equation}
\mathcal{L}_{\mathrm{SID}}(x)
=
\sum_{\ell=1}^{L^*}
w_\ell(x)
\operatorname{KL}\!\left(
\mathbf{q}_T^\ell\,\Vert\,\mathbf{p}_S^\ell
\right).
\end{equation}
Because the weights sum to one over $\mathcal{L}(x)$, SID loss does not increase mechanically with the shared-prefix length.

\subsection{Beam-Aware Ranking Distillation}
\label{sec:beam-aware-ranking-distillation}

Although SID loss handles imbalanced distillation difficulty across the SID hierarchy, it does not directly account for incorrect prefix pruning during beam search, where an item with a high final score may be removed because an intermediate prefix has a low cumulative score.
Since beam search determines candidate survival according to cumulative prefix rankings at each decoding step, mitigating this problem calls for explicitly optimizing these rankings during training.
Existing GR training methods mitigate this issue by optimizing target-item prefixes~\cite{yu2026apao,yang2026bear}.
In distillation, ranked teacher beams and their cumulative prefix scores provide additional supervision beyond the target-item supervision used in GR training.
Accordingly, we introduce a beam-aware ranking objective, named BEAM loss, which uses this beam-search ranking supervision to distill the teacher's preferences among candidate prefixes.

We use the teacher beam $b_{k^*}$ selected for SID loss as the positive beam.
To preserve the teacher's ranking preference while keeping the comparison difficult, we select the next lower-ranked beam as the hard negative:
\begin{equation}
k^-
=
k^*+1.
\end{equation}
When $k^*=K_{\mathrm{beam}}$, BEAM loss is not applied to this sample.
This pairwise design requires forced scoring of only two cached beam sequences, reducing the training overhead compared with listwise distillation over all teacher beams.
Comparisons with alternative negative-beam choices are reported in Appendix~\ref{app:rank-negative-selection}.

The student does not run beam search during training.
Instead, it uses forced scoring to compute the cumulative scores of the positive and negative beam prefixes:
\begin{equation}
S_{S,k}^{\ell}
=
\sum_{j=1}^{\ell}
\log p_S\!\left(z_{k,j}\mid x,\mathbf{z}_{k,<j}\right).
\end{equation}
Let $S_{T,k}^{\ell}$ denote the cumulative teacher score of the $k$-th beam prefix at SID level $\ell$.
BEAM loss is evaluated over the same set $\mathcal{L}(x)$ defined in Section~\ref{sec:hierarchy-aware-sid-distillation}, which contains the SID levels where the positive beam matches the target SID.
At each such level, we convert the cumulative scores of the positive and negative prefixes into teacher and student preference distributions:
\begin{equation}
\begin{aligned}
\boldsymbol{\pi}_T^{\ell}
&=
\operatorname{softmax}\!\left(
\frac{\left[S_{T,k^*}^{\ell},S_{T,k^-}^{\ell}\right]}{\ell\tau}
\right),\\
\boldsymbol{\pi}_S^{\ell}
&=
\operatorname{softmax}\!\left(
\frac{\left[S_{S,k^*}^{\ell},S_{S,k^-}^{\ell}\right]}{\ell\tau}
\right).
\end{aligned}
\end{equation}
where $\tau>0$ is the distillation temperature.
The gap between two cumulative prefix scores may increase at later SID levels as more token log-probabilities are accumulated.
Dividing the cumulative scores by $\ell$ converts them into average log-probabilities per SID level, reducing this length effect and making the preference distributions comparable across SID levels.
BEAM loss matches these preference distributions:
\begin{equation}
\mathcal{L}_{\mathrm{BEAM}}(x)
=
\frac{1}{|\mathcal{L}(x)|}
\sum_{\ell=1}^{L^*}
\tau^2
\operatorname{KL}\!\left(
\boldsymbol{\pi}_T^{\ell}\,\Vert\,\boldsymbol{\pi}_S^{\ell}
\right).
\end{equation}
This objective distills the teacher's relative preference between the two beam prefixes at each SID level in $\mathcal{L}(x)$.

SID loss distills the teacher's conditional SID distributions along the selected beam, whereas BEAM loss distills its ranking preference between candidate beam prefixes based on their cumulative scores.
The overall objective combines these two losses with hard supervision:
\begin{equation}
\mathcal{L}
=
\mathcal{L}_{\mathrm{hard}}
+
\lambda_{\mathrm{SID}}\mathcal{L}_{\mathrm{SID}}
+
\lambda_{\mathrm{BEAM}}\mathcal{L}_{\mathrm{BEAM}}.
\end{equation}
Here, $\mathcal{L}_{\mathrm{hard}}$ is the standard autoregressive loss on the target SID.
$\lambda_{\mathrm{SID}}$ and $\lambda_{\mathrm{BEAM}}$ are hyperparameters that trade off the contributions of the two distillation objectives.

\subsection{Discussion}
\paragraph{Core advantages of \method.}
\method addresses the GR-specific distillation challenges via two complementary perspectives:
(i) \emph{Token-level SID perspective:} Hierarchy-Aware SID Distillation learns a hierarchy-aware weight for the distillation loss at each SID position.
This adapts the distillation strength to the imbalanced distillation difficulty across the SID hierarchy.
(ii) \emph{Item-level beam search perspective:} Beam-Aware Ranking Distillation matches the teacher and student preference distributions constructed from the cumulative prefix scores of the positive and negative beams. It transfers the teacher's prefix-ranking preference to reduce incorrect prefix pruning during beam search.

\paragraph{Comparison with existing GR distillation methods.}
Several works have explored knowledge distillation for GR.
However, these methods do not systematically examine the teacher's advantages in hierarchical SID modeling and ignore cumulative prefix ranking during beam search.
SID-MLP~\cite{guo2026sidmlp} replaces the Transformer decoder with position-specific MLP heads, but this architectural difference introduces additional performance loss during distillation.
LOHRec~\cite{xie2025lohrec} distills codeword distributions and complete-item rankings, but ignores position-dependent distillation and intermediate prefix ranking, failing to fully leverage the teacher's advantages.
In contrast, \method transfers both advantages with two loss functions.
 \section{Experiments}
\label{sec:experiments}

\subsection{Experimental Setup}

\paragraph{Datasets.}
We evaluate \method on four real-world recommendation datasets: \textit{Amazon Beauty}, \textit{Amazon Toys}, \textit{Kuaishou Ad}, and \textit{Kuaishou Video}.
Dataset sources and preprocessing details are provided in Appendix~\ref{app:dataset-details}.
Table~\ref{tab:dataset-statistics} reports the statistics of the four datasets.

\begin{table}[t]
\centering
\normalsize
\setlength{\tabcolsep}{1pt}
\begin{tabular*}{\columnwidth}{@{\extracolsep{\fill}}lcccc@{}}
\toprule
\textbf{Statistic} & \textbf{Beauty} & \textbf{Toys} & \textbf{Video} & \textbf{Ad} \\
\midrule
\#Users & 18,041 & 15,001 & 156,245 & 121,056 \\
\#Items & 10,166 & 9,755 & 11,589,098 & 162,828 \\
\#Interactions & 147,383 & 125,166 & 75,671,142 & 4,273,723 \\
Density & 0.0804\% & 0.0855\% & 0.0042\% & 0.0217\% \\
\bottomrule
\end{tabular*}
\caption{Statistics of the datasets.}
\label{tab:dataset-statistics}
\end{table}
 
\begin{table*}[t]
\centering
\small
\setlength{\tabcolsep}{0.8pt}
\begin{tabular*}{\textwidth}{@{\extracolsep{\fill}}c|c|cccc|cccc|cccc|cccc@{}}
\toprule
\multirow{2}{*}{\textbf{Category}} & \multirow{2}{*}{\textbf{Method}} & \multicolumn{4}{c|}{\textbf{Beauty}} & \multicolumn{4}{c|}{\textbf{Toys}} & \multicolumn{4}{c|}{\textbf{Ad}} & \multicolumn{4}{c}{\textbf{Video}} \\
\cmidrule{3-18}
& & \textbf{R@5} & \textbf{R@10} & \textbf{N@5} & \textbf{N@10} & \textbf{R@5} & \textbf{R@10} & \textbf{N@5} & \textbf{N@10} & \textbf{R@16} & \textbf{R@32} & \textbf{P@16} & \textbf{P@32} & \textbf{R@16} & \textbf{R@32} & \textbf{P@16} & \textbf{P@32} \\
\midrule
\textbf{Student} & OneRec-1.7B & 4.14 & 6.06 & 2.89 & 3.52 & 4.19 & 6.01 & 2.79 & 3.37 & 4.81 & 7.39 & 14.56 & 21.24 & 2.45 & 2.74 & 15.07 & 16.95 \\
\textbf{Teacher} & OneRec-8B & 4.87 & 7.17 & 3.42 & 4.16 & 5.45 & 7.83 & 3.87 & 4.64 & 6.40 & 9.75 & 19.13 & 27.19 & 3.23 & 3.68 & 18.37 & 20.90 \\
\midrule
\multirow{6}{*}{\textbf{KD for RS}} & CD & 3.71 & 6.04 & 2.54 & 3.29 & 2.87 & 4.33 & 1.83 & 2.31 & 2.78 & 4.44 & 9.39 & 13.36 & 1.71 & 2.03 & 12.37 & 14.43 \\
& PCKD & 3.33 & 5.38 & 2.09 & 2.75 & 3.33 & 4.53 & 2.07 & 2.46 & 3.22 & 5.68 & 9.39 & 16.25 & 1.47 & 1.65 & 10.82 & 11.60 \\
& RCE-KD & 2.44 & 4.05 & 1.65 & 2.16 & 2.40 & 3.20 & 1.65 & 1.91 & 1.79 & 2.90 & 6.86 & 10.11 & 1.68 & 1.93 & 11.08 & 13.14 \\
& DLLM2Rec & 4.11 & 5.51 & 2.79 & 3.24 & 3.79 & 5.26 & 2.66 & 3.14 & 4.70 & 7.38 & 14.16 & 21.24 & 2.37 & 2.67 & 14.72 & 16.76 \\
& SLMRec & 2.31 & 3.34 & 1.58 & 1.92 & 2.38 & 2.92 & 1.55 & 1.73 & 2.83 & 4.48 & 8.89 & 13.51 & 1.87 & 2.09 & 12.33 & 13.95 \\
& C2KD & 4.04 & 5.79 & 2.80 & 3.36 & 3.81 & 5.28 & 2.68 & 3.15 & 4.02 & 6.48 & 12.38 & 18.83 & 2.41 & 2.71 & 15.03 & 17.03 \\
\midrule
\multirow{4}{*}{\textbf{KD for LLMs}} & KD & 4.21 & 5.83 & 2.93 & 3.45 & 4.28 & \underline{6.17} & 3.03 & 3.46 & \underline{5.12} & \underline{7.93} & \underline{15.67} & \underline{23.06} & 2.63 & 3.09 & 15.82 & 18.48 \\
& MiniLLM & \underline{4.37} & \underline{6.07} & \underline{3.03} & \underline{3.58} & \underline{4.58} & 5.61 & \underline{3.19} & \underline{3.71} & 5.00 & 7.75 & 15.28 & 22.29 & \underline{2.67} & 3.11 & \underline{15.85} & 18.38 \\
& SeqKD & 4.36 & 5.87 & \underline{3.03} & 3.54 & 3.87 & 5.43 & 2.68 & 3.18 & 4.21 & 5.92 & 10.84 & 15.66 & 2.58 & \underline{3.18} & 15.37 & \textbf{18.72} \\
& TVDKD & 3.55 & 4.67 & 2.52 & 2.89 & 2.44 & 3.17 & 1.70 & 1.94 & 4.10 & 6.37 & 12.65 & 18.93 & 2.28 & 2.59 & 14.81 & 16.95 \\
\midrule
\multirow{2}{*}{\textbf{KD for GR}} & LOHRec & 3.85 & 5.60 & 2.74 & 3.32 & 3.03 & 4.26 & 2.10 & 2.49 & 5.00 & 7.83 & 13.56 & 18.64 & 2.41 & 2.81 & \underline{15.85} & 18.29 \\
& SID-MLP & 3.76 & 5.36 & 2.64 & 3.47 & 2.96 & 3.92 & 1.96 & 2.41 & 4.14 & 7.19 & 13.49 & 17.68 & 2.21 & 2.62 & 15.01 & 16.85 \\
\midrule
\rowcolor{smartrow}
& \method & \textbf{4.40} & \textbf{6.20} & \textbf{3.04} & \textbf{3.63} & \textbf{4.64} & \textbf{6.31} & \textbf{3.23} & \textbf{3.76} & \textbf{5.15} & \textbf{8.02} & \textbf{15.69} & \textbf{23.15} & \textbf{2.72} & \textbf{3.22} & \textbf{15.97} & \underline{18.65} \\
& Impr.s (\%) & $+6.3$ & $+2.3$ & $+5.2$ & $+3.1$ & $+10.7$ & $+5.0$ & $+15.8$ & $+11.6$ & $+7.1$ & $+8.5$ & $+7.8$ & $+9.0$ & $+11.0$ & $+17.5$ & $+6.0$ & $+10.0$ \\
& Impr.b (\%) & $+0.7$ & $+2.1$ & $+0.3$ & $+1.4$ & $+1.3$ & $+2.3$ & $+1.3$ & $+1.3$ & $+0.6$ & $+1.1$ & $+0.1$ & $+0.4$ & $+1.9$ & $+1.3$ & $+0.8$ & $-0.4$ \\
\bottomrule
\end{tabular*}
\caption{Performance comparison on four recommendation datasets, where R, N, and P denote Recall, NDCG, and Pass, respectively. Bold and underlined values indicate the best and second-best student results, while Impr.s and Impr.b denote the relative improvements of \method over the original student and the best KD baseline, respectively.}
\label{tab:main-results}
\end{table*}
 
\paragraph{Baselines.}
We compare \method with three groups of baselines: (i) KD for RS, including CD~\cite{lee2019cd}, PCKD~\cite{zhu2025pckd}, RCE-KD~\cite{zhu2026rcekd}, DLLM2Rec~\cite{cui2024dllm2rec}, SLMRec~\cite{xu2025slmrec}, and C2KD~\cite{chen2025c2kd}; (ii) KD for LLMs, including KD~\cite{hinton2015distilling}, SeqKD~\cite{kim2016seqkd}, TVDKD~\cite{wen2023fdistill}, and MiniLLM~\cite{gu2024minillm}; and (iii) KD for GR, including LOHRec~\cite{xie2025lohrec} and SID-MLP~\cite{guo2026sidmlp}.
Readers can refer to Appendix~\ref{app:baseline-reproduction-details} for method descriptions and reproduction details.

\paragraph{Evaluation Metrics.}
Following previous works~\cite{zheng2024lcrec,zhou2025openonerec}, we employ Recall@$K$ and NDCG@$K$ for Amazon with $K=5,10$, and Pass@$K$ and Recall@$K$ for Kuaishou with $K=16,32$, as each Kuaishou sample contains multiple target items.
All reported metrics are higher-is-better.
Amazon and Kuaishou use beam widths 16 and 32, respectively.

\paragraph{Implementation Details.}
We perform offline distillation from OneRec-8B to OneRec-1.7B.
Following previous works~\cite{zhou2025openonerec}, we jointly train Beauty with Toys and Ad with Video, while evaluating each dataset separately.
We train the student for four epochs using AdamW with a learning rate of $5\times10^{-6}$ and a global batch size of 128.
By default, we set $K_{\mathrm{beam}}=16$, $\tau=\tau_{\mathrm{lev}}=1.0$, $\lambda_{\mathrm{SID}}=0.6$, and $\lambda_{\mathrm{BEAM}}=0.3$.
All loss-based baselines follow the same training and evaluation settings as \method.
Additional implementation details are provided in Appendix~\ref{app:baseline-reproduction}.

\begin{table}[t]
\centering
\normalsize
\setlength{\tabcolsep}{1.5pt}
\begin{tabular*}{\columnwidth}{@{\extracolsep{\fill}}lcccc@{}}
\toprule
\multirow{2}{*}{\textbf{Variant}} & \multicolumn{2}{c}{\textbf{Beauty}} & \multicolumn{2}{c}{\textbf{Toys}} \\
\cmidrule(lr){2-3}\cmidrule(lr){4-5}
& \textbf{N@5} & \textbf{N@10} & \textbf{N@5} & \textbf{N@10} \\
\midrule
Base & 2.890 & 3.520 & 2.790 & 3.370 \\
w/o SID loss & 2.893 & 3.527 & \underline{3.146} & \underline{3.727} \\
w/o BEAM loss & \underline{2.910} & \underline{3.594} & 3.040 & 3.673 \\
w/o Longest Prefix & 2.867 & 3.461 & 2.785 & 3.291 \\
w/o Overlapping Positions & 2.877 & 3.481 & 2.772 & 3.273 \\
Full \method & \textbf{3.044} & \textbf{3.626} & \textbf{3.226} & \textbf{3.761} \\
\bottomrule
\end{tabular*}
\caption{Component ablations on Amazon; bold and underlined values denote the best and second-best results.}
\label{tab:component-ablation}
\end{table}
 
\subsection{Performance Comparison}

Table~\ref{tab:main-results} compares \method with the original student, the teacher, and three groups of knowledge-distillation strategies on four recommendation datasets.

\paragraph{Overall Performance.}
Compared with the original OneRec-1.7B student, \method improves all 16 metrics by 2.3\%--17.5\%, with an average gain of 8.6\%.
It achieves the best result among all baselines on 15 of the 16 metrics across the four datasets.
These consistent gains demonstrate that \method effectively transfers the teacher's advantages to the lightweight student across both Amazon and Kuaishou datasets.

\paragraph{Compared with KD for RS.}
\method outperforms all six KD-for-RS baselines on every metric, with a 12.1\% average gain.
Notably, all six baselines underperform the original student.
This could be attributed to complete-item supervision, which cannot provide exact token-level signals and may fail to distinguish items that share SID prefixes.

\paragraph{Compared with KD for LLMs.}
\method achieves a 1.0\% average gain over the best KD-for-LLMs results across 16 metrics.
KD and MiniLLM use position-agnostic token-level supervision, whereas SeqKD transfers SID sequences without modeling SID-level difficulty.
TVDKD's bounded objective may provide limited correction for underestimated teacher-preferred tokens, allowing errors to propagate.
Together, these limitations make generic language-model distillation less consistent across datasets than \method.

\paragraph{Compared with KD for GR.}
\method outperforms LOHRec and SID-MLP on every metric by 20.4\% and 27.3\% on average, respectively.
SID-MLP replaces the autoregressive decoder with MLP heads, resulting in substantial performance degradation.
Although LOHRec transfers teacher codeword distributions and complete-candidate rankings, it does not explicitly capture the teacher's advantages.
These results highlight the importance of jointly modeling hierarchical SID generation and beam-search ranking.

\subsection{Ablation Study}

To assess the contribution of each design choice, Table~\ref{tab:component-ablation} reports four ablations on Amazon Beauty and Toys.
\textit{w/o SID loss} and \textit{w/o BEAM loss} remove the corresponding objectives.
\textit{w/o Longest Prefix} selects the highest-scoring teacher beam instead of using longest-prefix selection, whereas \textit{w/o Overlapping Positions} distills all positions on the selected beam.

Removing SID loss and BEAM loss yields average relative declines of 2.8\% and 3.3\%, respectively.
Nevertheless, \textit{w/o SID loss} ranks second on Toys, whereas \textit{w/o BEAM loss} ranks second on Beauty, suggesting that Toys relies more on beam-search ranking and Beauty more on hierarchical SID modeling.
Replacing longest-prefix selection or removing the overlap mask causes an average relative decline of 9.1\% in both cases.
These results show that teacher beams selected for distillation should remain compatible with the target SID and that distilling tokens inconsistent with the ground truth introduces harmful supervision.

\subsection{Efficiency Study}

Table~\ref{tab:beam-distillation-efficiency} compares performance and total distillation time under different teacher beam widths.
Increasing the teacher beam width yields a substantial 8.5\% average gain, while requiring only $1.53\times$ as much distillation time.
\begin{table}[t]
\centering
\normalsize
\setlength{\tabcolsep}{1.5pt}
\begin{tabular*}{\columnwidth}{@{\extracolsep{\fill}}cccccc@{}}
\toprule
\multirow{2}{*}{\textbf{$K_{\mathrm{beam}}$}}
& \multicolumn{2}{c}{\textbf{Beauty}}
& \multicolumn{2}{c}{\textbf{Toys}}
& \multirow{2}{*}{\textbf{\shortstack{Time\\(h)}}} \\
\cmidrule(lr){2-3}\cmidrule(lr){4-5}
& \textbf{R@5} & \textbf{R@10}
& \textbf{R@5} & \textbf{R@10} & \\
\midrule
2  & 3.991 & 5.499 & 4.346 & 6.046 & 2.06 \\
4  & 4.107 & 5.698 & 4.553 & 6.206 & 2.22 \\
8  & \underline{4.340} & \underline{6.086} & \underline{4.580} & \underline{6.293} & 2.57 \\
16 & \textbf{4.396} & \textbf{6.203} & \textbf{4.640} & \textbf{6.306} & 3.16 \\
\bottomrule
\end{tabular*}
\caption{Recall performance and total training time, including cache construction and training on Amazon.}
\label{tab:beam-distillation-efficiency}
\end{table}
 Table~\ref{tab:inference-speedup} compares the inference time of OneRec-1.7B and OneRec-8B across four datasets.
Because \method leaves the student architecture unchanged, the distilled model retains OneRec-1.7B's efficiency and is $1.92\times$--$2.89\times$ faster than OneRec-8B.
We additionally report a comparison of the pairwise BEAM objective with listwise supervision in Appendix~\ref{app:ranking-supervision-scope}.
\looseness=-1
\begin{table}[t]
\centering
\normalsize
\setlength{\tabcolsep}{1.5pt}
\begin{tabular*}{\columnwidth}{@{\extracolsep{\fill}}ccccc@{}}
\toprule
\textbf{Model} & \textbf{Beauty} & \textbf{Toys} & \textbf{Ad} & \textbf{Video} \\
\midrule
OneRec-1.7B & 0.86 & 0.71 & 0.31 & 1.05 \\
OneRec-8B & 2.49 & 1.97 & 0.59 & 2.09 \\
\cmidrule{1-5}
Gain & $2.89\times$ & $2.77\times$ & $1.92\times$ & $1.99\times$ \\
\bottomrule
\end{tabular*}
\caption{Inference time (h) and OneRec-1.7B speedup over OneRec-8B at beam widths 16 (Amazon) and 32 (Kuaishou).}
\label{tab:inference-speedup}
\end{table}
 
\subsection{Case Study}

\begin{figure}[t]
  \centering
  \includegraphics[width=\columnwidth]{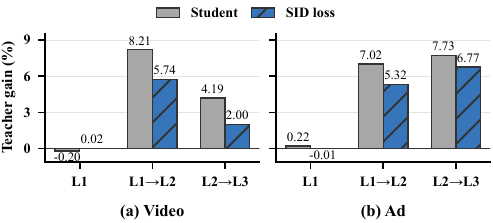}
  \caption{Teacher gains over the original student and the student trained with SID loss in target-item prefix retention.}
  \label{fig:challenge-analysis}
\end{figure}

Figure~\ref{fig:challenge-analysis} reports the teacher's prefix-retention gains over the original student and the student trained with SID loss across SID transitions.
SID loss reduces the mean absolute gap to the teacher by 28.0\% across six transitions and yields a smaller gap than the original student at five of them, demonstrating that it effectively addresses varying distillation difficulty across SID levels.
Figure~\ref{fig:positive-beam-pruning}(b) reports the mean prefix rank of each model's global Top-1 item.
BEAM loss reduces this rank by 46.5\% on average relative to the original student and surpasses the teacher on Ad, demonstrating that it effectively improves prefix ranking during beam search.
\looseness=-1
 \section{Related Work}
\label{sec:related-work}

\paragraph{Generative Recommendation.}
Generative recommendation (GR) is an emerging paradigm that autoregressively generates semantic IDs (SIDs), combining item semantics with scaling potential~\cite{rajput2023tiger,zhou2025openonerec}.
GR scales through larger models, more training data, and larger systems~\cite{zhai2024hstu,han2025mtgr,zhou2025openonerec,kong2025minionerec}.
For instance, OneRec~\cite{zhou2025openonerec} evaluates scaling with larger models and more compute.
Larger GR models improve recommendation performance but incur high computation and latency, making efficient GR an urgent research problem~\cite{lin2025atspeed,guo2026sidmlp}.

\paragraph{Knowledge Distillation for Recommendation.}
Knowledge distillation (KD) is a model-compression technique that transfers knowledge from a large teacher to a compact student~\cite{hinton2015distilling}, and has been widely applied to recommendation to reduce inference latency~\cite{tang2018rd,lee2019cd}.
Traditional recommendation distillation transfers item scores or rankings~\cite{tang2018rd,lee2019cd}, intermediate representations~\cite{zhu2025pckd,zhu2025freqd}, or collaborative structures~\cite{kang2021topology}.
However, these traditional methods predict the next item ID based on item logits, different from GR's hierarchical SID tokens and beam-search decoding, making them difficult to be applied into GR directly.

KD is also used to reduce inference latency in LLM-based recommendation~\cite{xu2025slmrec,ramos2024peapod}. For example, DLLM2Rec~\cite{cui2024dllm2rec} distills ranking and representation knowledge, while C2KD~\cite{chen2025c2kd} distills selected teacher's layers knowledge.
However, the item description in LLM are long, variable-length language token sequences, unlike GR's short, coarse-to-fine SIDs, and therefore remain unsuitable for GR.

While recent works have preliminarily explored KD for GR, they remain at a surface level, lacking in-depth designs tailored to GR mechanisms.
SID-MLP~\cite{guo2026sidmlp} replaces the decoder with position-specific MLP heads, causing performance loss due to the architectural difference.
LOHRec~\cite{xie2025lohrec} distills codeword distributions and item rankings but ignores position-dependent distillation and prefix ranking.
\method transfers teacher's advantages in hierarchical SID modeling and beam-search ranking through two losses, enabling GR-specific distillation.
 \section{Conclusion}
\label{sec:conclusion}

In this paper, we reveal the key challenges of employing knowledge distillation in generative recommendation: imbalanced distillation difficulty across the SID hierarchy and incorrect prefix pruning during beam search. To overcome these limitations, we propose \method, a novel GR distillation framework that utilizes Hierarchy-Aware SID Distillation to adapt the distillation strength across SID positions and uses Beam-Aware Ranking Distillation to transfer the teacher's ranking preferences during beam search.
Extensive experiments on four benchmark datasets demonstrate the effectiveness and efficiency of our method.
 
\FloatBarrier
\newpage
\bibliography{references}

\clearpage
\appendix
\section{Experimental Details}
\label{app:baseline-reproduction}

\subsection{Dataset Details}
\label{app:dataset-details}

We evaluate \method on four real-world recommendation datasets from two sources; their statistics are reported in Table~\ref{tab:dataset-statistics} of the main paper.

\begin{itemize}[leftmargin=10pt,noitemsep,topsep=0pt]
    \item \textbf{Amazon Reviews}\footnote{\url{https://cseweb.ucsd.edu/~jmcauley/datasets/amazon/links.html}}~\cite{he2016ups} is a widely used e-commerce recommendation benchmark containing timestamped user--item review interactions. We use its \textit{Beauty} and \textit{Toys} subsets. Following prior generative recommendation work~\cite{zheng2024lcrec}, we apply 5-core filtering, such that every retained user and item has at least five interactions. We then order each user's interactions chronologically and adopt leave-one-out evaluation, treating the last item as the prediction target and the preceding items as the interaction history. Each item is represented by a four-level SID comprising three semantic levels followed by a collision-resolution position.
    \item \textbf{OpenOneRec-RecIF}~\cite{zhou2025openonerec}\footnote{\url{https://huggingface.co/datasets/OpenOneRec/OpenOneRec-RecIF}} is a multi-domain recommendation benchmark constructed from anonymized Kuaishou user behaviors. Following previous work, we use its \textit{Ad} and \textit{Video} recommendation tasks. The former predicts clicked ads from users' historical ad clicks and long-view video behaviors, whereas the latter predicts target videos from chronological video interaction histories. Each sample contains multiple target items. We retain the provided user-level split and temporal evaluation protocol. Each video or ad item is represented by a three-level SID.
\end{itemize}

\subsection{\method Implementation}
\label{app:smartgr-implementation}

We perform offline distillation from OneRec-8B to OneRec-1.7B.
For Kuaishou, we directly use the released teacher and student checkpoints.
For Amazon, following the OpenOneRec technical report, we append a finite scalar quantization (FSQ)-based collision-resolution position as the fourth SID level and fine-tune OneRec-8B and OneRec-1.7B as the teacher and base student, respectively.
To construct the offline teacher cache, we set the beam width $K_{\mathrm{beam}}$ to 16 and retain the Top-$K_{\mathrm{tok}}$ teacher token distribution at each valid SID position, with $K_{\mathrm{tok}}=16$.
For hierarchy-aware SID weighting, we set $\theta_{\max}=4.0$ and $\tau_{\mathrm{lev}}=1.0$, and use minimum shared-prefix lengths of two and three for Amazon and Kuaishou, respectively.
For BEAM loss, the distillation temperature $\tau$ is set to 1.0.
Finally, we set the objective coefficients to $\lambda_{\mathrm{SID}}=0.6$ and $\lambda_{\mathrm{BEAM}}=0.3$.

\subsection{Training and Evaluation Protocol}
\label{app:training-evaluation-protocol}

\paragraph{Training and evaluation settings.}
Unless otherwise specified, all methods use the same dataset splits, OneRec-1.7B initialization, ground-truth SID supervision, and evaluation protocol; method-specific sampling and optimization settings are detailed in Appendix~\ref{app:baseline-reproduction-details}.
Beauty and Toys are jointly trained, as are Ad and Video.
We train \method for four epochs using AdamW with a learning rate of $5\times10^{-6}$, a weight decay of $0.1$, and a global batch size of 128.
We do not apply early stopping and select the checkpoint with the lowest validation hard loss.
We implement \method using PyTorch 2.6.0, Transformers 4.51.1, DeepSpeed 0.16.5 with ZeRO Stage 2, BF16, and FlashAttention 2 on eight NVIDIA A100 GPUs.
For autoregressive inference, we use beam widths of 16 for Amazon and 32 for Kuaishou.
Tables~1 and~7 of the main paper report throughput-oriented total wall-clock inference time under different vLLM configurations.
Within each table, the teacher and student use the same configuration, so their times can be compared.
The absolute times cannot be compared between the two tables.

\paragraph{Evaluation metrics.}
\label{app:evaluation-metrics}
We use Recall@$K$ and NDCG@$K$ for Amazon, and Recall@$K$ and Pass@$K$ for Kuaishou, with the cutoffs specified in the main paper.
Following the OpenOneRec evaluator, Kuaishou metrics are computed at the physical item ID (PID) level after mapping generated SIDs back to items.
Let $\mathcal{D}$ denote the evaluation set, $\mathcal{R}_K(x)$ the set of the top-$K$ retrieved PIDs for context $x$, and $\mathcal{Y}(x)$ its set of target PIDs.
Because each Kuaishou sample contains multiple target items, Pass@$K$ measures the fraction of examples for which at least one target is retrieved:
\begin{equation}
\operatorname{Pass@K}
=
\frac{1}{|\mathcal{D}|}
\sum_{x\in\mathcal{D}}
\mathbb{I}[\mathcal{R}_K(x)\cap\mathcal{Y}(x)\neq\varnothing].
\label{eq:pass-at-k}
\end{equation}

\subsection{Baseline Reproduction}
\label{app:baseline-reproduction-details}

We adapt all baselines to the same OneRec-8B to OneRec-1.7B distillation setting used by \method.
All loss-based baselines start from the hard-supervised OneRec-1.7B checkpoint and retain ground-truth SID supervision unless stated otherwise.
We follow the original formulation of each baseline and tune method-specific hyperparameters on the validation set.

\subsubsection{Knowledge Distillation for Recommendation Systems}

This group includes item-level recommendation KD methods (CD, PCKD, and RCE-KD) and KD methods for LLM-based sequential recommenders (DLLM2Rec, SLMRec, and C2KD).
To adapt the first three methods to GR, we treat each complete SID sequence as one candidate item and use its autoregressive sequence log-likelihood as the item score.
These methods therefore operate on complete candidates without longest-common-prefix filtering.

\begin{itemize}[leftmargin=10pt,noitemsep,topsep=0pt]
    \item \textbf{CD}~\cite{lee2019cd} uses student-guided rank-aware sampling over complete candidates and transfers soft binary targets derived from cached teacher sequence scores. We sample half of the candidates and set the distillation weight to 0.5.
    \item \textbf{PCKD}~\cite{zhu2025pckd} reproduces the pairwise PCKD-P variant by projecting the student user representation and mean-pooled SID item embeddings to the teacher space. Two candidates are sampled with temperature 10. We set the feature-matching weight, pairwise preference-preservation weight, and overall PCKD loss weight to 1.0, 0.005, and 1.0, respectively.
    \item \textbf{RCE-KD}~\cite{zhu2026rcekd} constructs teacher and student top sets from complete-item sequence scores and adaptively emphasizes teacher-preferred items ranked poorly by the student. We use sampling temperature 10, overlap coefficient 5, and loss weight 5.0.
    \item \textbf{DLLM2Rec}~\cite{cui2024dllm2rec} assigns importance weights to complete teacher-beam SIDs based on teacher rank, target-prefix confidence, and teacher--student Top-10 consistency. It additionally aligns mean-pooled student SID states with projected teacher SID embeddings, using up to four candidates on Amazon and ten on Kuaishou.
    \item \textbf{SLMRec}~\cite{xu2025slmrec} applies block-wise layer distillation at target-SID positions while retaining the OneRec-1.7B architecture. We align teacher layers $\{9,18,27,36\}$ with student layers $\{7,14,21,28\}$ through learnable projectors and retain the original feature-matching and auxiliary SID-prediction objectives.
    \item \textbf{C2KD}~\cite{chen2025c2kd} dynamically routes the last two student layers to teacher layers $\{18,24,30,36\}$ for feature imitation. The projected final student state is also passed through the frozen teacher output head to match the teacher's Top-64 token distribution.
\end{itemize}

\subsubsection{Knowledge Distillation for Large Language Models}

We restrict token-level supervision to causal positions that predict SID symbols.
For KD, MiniLLM, and TVDKD, both teacher and student distributions are renormalized over the cached Top-16 teacher-token support available at each supervised position.
SeqKD instead learns from complete teacher-generated SID sequences.

\begin{itemize}[leftmargin=10pt,noitemsep,topsep=0pt]
    \item \textbf{KD}~\cite{hinton2015distilling} follows the cached teacher beam with the longest target-compatible prefix and minimizes $\operatorname{KL}(p_{\mathrm{T}}\Vert p_{\mathrm{S}})$ over the SID positions within this shared prefix.
    \item \textbf{MiniLLM}~\cite{gu2024minillm} uses the same selected prefixes and cached support as KD but reverses the divergence direction to $\operatorname{KL}(p_{\mathrm{S}}\Vert p_{\mathrm{T}})$, yielding an offline reverse-KL adaptation under our shared cache-based protocol.
    \item \textbf{SeqKD}~\cite{kim2016seqkd} uses the teacher's two highest-ranked complete SID sequences as pseudo targets and averages their autoregressive sequence cross-entropies. We combine this term with hard supervision as $\mathcal{L}_{\mathrm{CE}}+0.6\mathcal{L}_{\mathrm{SeqKD}}$.
    \item \textbf{TVDKD}~\cite{wen2023fdistill} replaces KL matching with total variation distance at positions teacher-forced by the ground-truth SID. Its objective is $0.3\mathcal{L}_{\mathrm{CE}}+0.7\mathcal{L}_{\mathrm{TVD}}$.
\end{itemize}

\subsubsection{Knowledge Distillation for Generative Recommendation}

We compare with two KD methods developed specifically for SID-based generative recommendation.

\begin{itemize}[leftmargin=10pt,noitemsep,topsep=0pt]
    \item \textbf{LOHRec}~\cite{xie2025lohrec} uses cached OneRec-8B codeword distributions and ordered complete SID candidates produced by teacher beam search to supervise OneRec-1.7B, while retaining constrained decoding over valid quantized SIDs. Its ranking objective operates on complete candidates rather than cumulative prefix rankings at each SID depth.
    \item \textbf{SID-MLP}~\cite{guo2026sidmlp} uses OneRec-1.7B as the backbone but replaces its autoregressive Transformer decoder with prefix-conditioned, position-specific MLP heads. These heads are trained with ground-truth SIDs and cached OneRec-8B logits at the corresponding SID positions.
\end{itemize}

Both baselines use the same datasets and evaluation protocol as \method.
LOHRec preserves the autoregressive OneRec student architecture, whereas SID-MLP follows its original non-autoregressive decoder design.
 \section{Additional Ablation Studies}
\label{app:amazon-additional-ablation}

This section complements the main experiments by examining hyperparameter sensitivity and alternative designs for SID and BEAM distillation, extending the component ablations to Kuaishou, and analyzing computational and storage complexity.
\FloatBarrier

\subsection{Hyperparameter Sensitivity}
\label{app:loss-weight-sensitivity}

\begin{figure*}[!t]
    \centering
    \includegraphics[width=\textwidth]{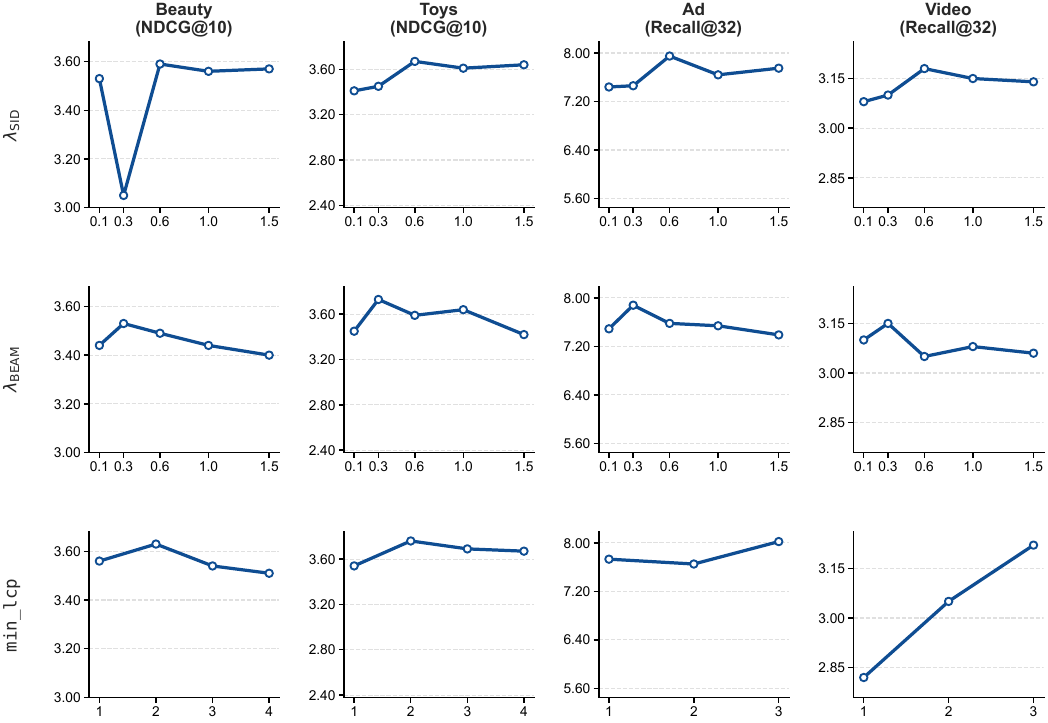}
    \caption{Hyperparameter sensitivity on Beauty, Toys, Ad, and Video. Each row varies one hyperparameter, and the columns report NDCG@10 on Amazon and Recall@32 on Kuaishou.}
    \label{fig:hyperparameter-sensitivity}
\end{figure*}

Figure~\ref{fig:hyperparameter-sensitivity} reports the sensitivity of \method to $\lambda_{\mathrm{SID}}$, $\lambda_{\mathrm{BEAM}}$, and \texttt{min\_lcp}.
Each loss-weight sweep activates only the corresponding distillation loss while keeping the remaining settings fixed, whereas the \texttt{min\_lcp} sweep uses both losses with their default coefficients.
The settings $\lambda_{\mathrm{SID}}=0.6$ and $\lambda_{\mathrm{BEAM}}=0.3$ achieve the strongest performance across the four datasets.
More importantly, performance remains close to its best value over most neighboring coefficients, showing that \method is stable under moderate changes to the loss weights rather than relying on a narrowly tuned configuration.

The \texttt{min\_lcp} sweep examines the trade-off between the reliability and coverage of teacher supervision.
A permissive threshold may retain teacher beams with insufficient agreement with the target SID, introducing distillation signals that interfere with ground-truth learning.
Conversely, an overly strict threshold can discard otherwise useful teacher supervision.
The different preferred thresholds on Amazon and Kuaishou reflect this dataset-dependent trade-off.
Overall, the results confirm the importance of teacher--target consistency and support longest-prefix beam selection together with overlap-restricted supervision.

\subsection{SID-Level Weighting}
\label{app:sid-weighting-ablation}

To assess both the monotonicity constraint and the choice of mapping function, Table~\ref{tab:sid-weighting-ablation} compares Base with uniform, per-level, sigmoid, exponential, and $\tanh$ weighting.
All weighting variants use SID loss only, with BEAM loss disabled and all other settings unchanged.
Base excludes SID distillation, whereas Uniform (KD) reproduces the KD baseline reported in the main paper by assigning $w_\ell(x)=1/L^*$ to every valid SID level.
Sigmoid, exponential, and $\tanh$ enforce weights that increase with SID depth, whereas per-level weights learn each level independently without this constraint.
In the sigmoid and exponential variants, the corresponding activation directly replaces $\tanh$ in $f_\theta$, while the remaining weight normalization and SID-loss formulation are unchanged.

Per-level weights assign a separate trainable logit $\alpha_\ell\in\mathbb{R}$ to each absolute SID level and normalize the logits over the valid shared prefix:
\begin{equation}
w_\ell^{\mathrm{lvl}}(x)
=
\frac{\exp(\alpha_\ell)}
{\sum_{j=1}^{L^*}\exp(\alpha_j)},
\qquad
\ell\in\mathcal{L}(x).
\label{eq:per-level-weighting}
\end{equation}
It then replaces $w_\ell(x)$ in SID loss with $w_\ell^{\mathrm{lvl}}(x)$:
\begin{equation}
\mathcal{L}_{\mathrm{SID}}^{\mathrm{lvl}}(x)
=
\sum_{\ell=1}^{L^*}
w_\ell^{\mathrm{lvl}}(x)
\operatorname{KL}\!\left(
\mathbf{q}_T^\ell\,\Vert\,\mathbf{p}_S^\ell
\right).
\label{eq:per-level-sid-loss}
\end{equation}
All $\alpha_\ell$ are initialized to zero, so the variant starts from uniform weighting.
The logits are shared across examples and learned jointly with each student model.

\begin{table}[t]
\centering
\normalsize
\setlength{\tabcolsep}{1.5pt}
\begin{tabular*}{\columnwidth}{@{\extracolsep{\fill}}lcccc@{}}
\toprule
\multirow{2}{*}{\textbf{Weighting function}} & \multicolumn{2}{c}{\textbf{Beauty}} & \multicolumn{2}{c}{\textbf{Toys}} \\
\cmidrule(lr){2-3}\cmidrule(lr){4-5}
& \textbf{N@5} & \textbf{N@10} & \textbf{N@5} & \textbf{N@10} \\
\midrule
Base & 2.890 & 3.520 & 2.790 & 3.370 \\
Uniform (KD) & 2.903 & 3.450 & \underline{3.030} & 3.460 \\
Per-level weights & 2.898 & \underline{3.521} & 2.752 & 3.229 \\
Sigmoid & \underline{2.906} & \underline{3.521} & 3.020 & \underline{3.657} \\
Exponential & 2.905 & 3.509 & 3.015 & 3.605 \\
$\tanh$ & \textbf{2.910} & \textbf{3.594} & \textbf{3.040} & \textbf{3.673} \\
\bottomrule
\end{tabular*}
\caption{Comparison of SID-level weighting schemes on Amazon using SID loss only; bold and underlined values denote the best and second-best results.}
\label{tab:sid-weighting-ablation}
\end{table}

The $\tanh$ mapping performs best overall and consistently improves over Base on both datasets.
Uniform weighting is less effective overall, supporting the use of different distillation weights across SID levels rather than averaging their supervision.
Per-level weights provide no consistent gain and underperform the monotonically increasing mappings overall, particularly on Toys.
Their weaker performance suggests that additional flexibility alone does not reliably recover the relative importance of SID levels; explicitly encoding the coarse-to-fine depth ordering provides a more effective structural prior.
Among the monotone alternatives, the results further justify $\tanh$ as an effective and compact parameterization of this depth ordering.

\subsection{BEAM Negative Selection}
\label{app:rank-negative-selection}

An effective negative beam should satisfy two requirements: it should be ranked below the positive beam by the teacher, while remaining sufficiently difficult for the student to distinguish.
Since teacher beams are sorted by their final sequence scores, the next lower-ranked beam \(k^{-}=k^*+1\) is the closest admissible candidate in the teacher's ranking.
It therefore preserves the teacher's pairwise preference and is more likely to provide an informative rank-local contrast.
In comparison, random selection provides inconsistent negative difficulty.
Using the first-ranked beam as the negative does not faithfully represent the teacher's beam-search preference because it places a teacher-preferred candidate below the selected positive in the constructed pair.
The last-ranked beam preserves the ordering direction but is typically too easy to distinguish, resulting in a weak contrast.

When \(k^*=K_{\mathrm{beam}}\), no valid lower-ranked beam is available, and BEAM loss is therefore omitted for that example.
We do not use the preceding beam as a fallback because the teacher ranks it above the selected positive, which would reverse the intended pairwise relation.
This boundary case affects fewer than \(1.2\%\) of cached examples on every dataset.
Such examples are not removed from training: they continue to receive ground-truth supervision, while SID loss remains active whenever the selected teacher beam has a valid shared prefix with the target.

Table~\ref{tab:rank-design-ablation} compares different negative-beam selection strategies under the same BEAM-only setting.
The next lower-ranked strategy performs consistently best across both Amazon datasets.
The first-ranked beam fails to provide a faithful teacher-ordering signal for the selected positive, whereas the last-ranked beam provides insufficient contrast; random selection does not control either property consistently.
The comparison therefore shows that effective BEAM supervision requires both teacher-consistent ordering and a sufficiently difficult negative.
These results support selecting the nearest lower-ranked beam as a simple and reliable way to transfer the teacher's local ranking preference.

\begin{table}[!htbp]
\centering
\normalsize
\setlength{\tabcolsep}{1.5pt}
\begin{tabular*}{\columnwidth}{@{\extracolsep{\fill}}lcccc@{}}
\toprule
\multirow{2}{*}{\textbf{Negative beam}} & \multicolumn{2}{c}{\textbf{Beauty}} & \multicolumn{2}{c}{\textbf{Toys}} \\
\cmidrule(lr){2-3}\cmidrule(lr){4-5}
& \textbf{N@5} & \textbf{N@10} & \textbf{N@5} & \textbf{N@10} \\
\midrule
Base & \underline{2.890} & \underline{3.520} & 2.790 & 3.370 \\
Random & 2.877 & 3.452 & \underline{3.129} & 3.692 \\
First & 2.873 & 3.454 & 3.127 & \underline{3.693} \\
Last & 2.650 & 3.093 & 2.873 & 3.414 \\
Next lower-ranked & \textbf{2.893} & \textbf{3.527} & \textbf{3.146} & \textbf{3.727} \\
\bottomrule
\end{tabular*}
\caption{Comparison of random, first, last, and next-lower-ranked negative-beam selection on Amazon; bold and underlined values denote the best and second-best results.}
\label{tab:rank-design-ablation}
\end{table}
 
\subsection{Ranking-Supervision Scope}
\label{app:ranking-supervision-scope}

BEAM loss transfers the teacher's ranking preference through a local comparison between the selected positive beam and its next lower-ranked candidate.
To examine whether a broader supervision scope provides additional ranking information, we compare the proposed pairwise objective with Top-4 and Top-8 listwise alternatives.
The listwise variants extend ranking supervision to multiple retained teacher beams, while using the same teacher cache, student model, and training and evaluation settings.

\begin{table}[!htbp]
\centering
\normalsize
\setlength{\tabcolsep}{1.5pt}
\begin{tabular*}{\columnwidth}{@{\extracolsep{\fill}}lccc@{}}
\toprule
\textbf{Metric} & \textbf{Pairwise} & \textbf{\shortstack{Top-4\\listwise}} & \textbf{\shortstack{Top-8\\listwise}} \\
\midrule
Beauty N@10 & \textbf{3.626} & 3.353 & \underline{3.368} \\
Toys N@10 & \textbf{3.761} & 3.396 & \underline{3.466} \\
Training time (h) & 1.35 & 2.21 & 2.55 \\
\bottomrule
\end{tabular*}
\caption{Recommendation performance and training time for pairwise, Top-4 listwise, and Top-8 listwise supervision on Beauty and Toys. N@10 denotes NDCG@10; bold and underlined values denote the best and second-best recommendation results.}
\label{tab:rank-supervision-efficiency}
\end{table}
 
As shown in Table~\ref{tab:rank-supervision-efficiency}, pairwise supervision consistently outperforms both listwise alternatives on the two Amazon datasets while reducing training time by 39\%--47\%.
Expanding the supervision scope from four to eight beams increases the training cost but does not close the performance gap with the pairwise objective.

These results suggest that using more teacher beams does not necessarily provide more effective ranking supervision.
Beam survival is determined by competition among nearby candidates, making the immediately lower-ranked beam the most relevant contrast for the selected positive.
In comparison, listwise supervision introduces additional and often easier comparisons with more distant beams, which can dilute the local ranking signal while requiring forced scoring of more sequences.
Therefore, the proposed pairwise objective is not only more efficient but also better focused on the teacher preference most relevant to beam pruning.
Together with the negative-selection results in Section~\ref{app:rank-negative-selection}, this comparison supports transferring the teacher's ranking knowledge through a local adjacent-beam comparison.

\subsection{Kuaishou Component Ablation}
\label{app:kuaishou-additional-ablation}

To examine whether the component-level conclusions generalize beyond Amazon, Table~\ref{tab:kuaishou-component-ablation} repeats the ablation study on Kuaishou Ad and Video.
Each variant changes only the indicated component while retaining the same student initialization, training configuration, and evaluation protocol as full \method.
Specifically, \textit{w/o SID loss} and \textit{w/o BEAM loss} remove the corresponding distillation objectives.
\textit{w/o Longest Prefix} selects the highest-scoring teacher beam instead of the beam with the longest target-compatible prefix, whereas \textit{w/o Overlapping Positions} applies distillation to all SID positions of the selected beam.

\begin{center}
\normalsize
\setlength{\tabcolsep}{1.5pt}
\begin{tabular*}{\columnwidth}{@{\extracolsep{\fill}}lcccc@{}}
\toprule
\multirow{2}{*}{\textbf{Variant}} & \multicolumn{2}{c}{\textbf{Ad}} & \multicolumn{2}{c}{\textbf{Video}} \\
\cmidrule(lr){2-3}\cmidrule(lr){4-5}
& \textbf{P@32} & \textbf{R@32} & \textbf{P@32} & \textbf{R@32} \\
\midrule
Base & 21.24 & 7.39 & 16.95 & 2.74 \\
w/o SID loss & 22.14 & 7.88 & 17.77 & 3.15 \\
w/o BEAM loss & \underline{22.28} & \underline{7.95} & \underline{18.21} & \underline{3.18} \\
w/o Longest Prefix & 16.74 & 6.01 & 15.41 & 2.39 \\
w/o Overlapping Positions & 17.04 & 6.11 & 14.53 & 2.25 \\
Full \method & \textbf{23.15} & \textbf{8.02} & \textbf{18.65} & \textbf{3.22} \\
\bottomrule
\end{tabular*}
\captionof{table}{Component ablations on Kuaishou Ad and Video. P@32 and R@32 denote Pass@32 and Recall@32, respectively; bold and underlined values denote the best and second-best results.}
\label{tab:kuaishou-component-ablation}
\end{center}
 
Full \method consistently achieves the strongest performance across both datasets.
Removing either SID loss or BEAM loss degrades the results, confirming that hierarchy-aware SID supervision and beam-ranking supervision provide complementary benefits.
More substantial degradation occurs when longest-prefix selection or overlap-restricted supervision is removed.
In these variants, teacher information that is insufficiently aligned with the target SID is introduced into distillation and can interfere with ground-truth learning, substantially weakening the benefit of teacher supervision.

The same pattern is observed in the Amazon ablation reported in the main paper: both objectives contribute to the final performance, while maintaining teacher--target compatibility is critical for effective distillation.
The consistent conclusions across Amazon and Kuaishou show that the proposed components are effective under different SID lengths and recommendation domains.
 
\subsection{Computational and Storage Complexity}
\label{app:computational-storage-complexity}

Let $N$ denote the number of training examples, $L$ the SID length, $K_{\mathrm{beam}}$ the number of cached teacher beams, $K_{\mathrm{tok}}$ the sparse token-support size, and $V$ the SID-token vocabulary size.
Compared with online distillation, the offline teacher cache eliminates teacher forward passes during student training.
Compared with dense-distribution caching, retaining only the Top-$K_{\mathrm{tok}}$ teacher tokens reduces the cache complexity from $O(NK_{\mathrm{beam}}LV)$ to $O(NK_{\mathrm{beam}}L(1+K_{\mathrm{tok}}))$, while SID loss requires only $O(L^*K_{\mathrm{tok}})$ sparse distribution operations for a shared prefix of length $L^*$.
Compared with listwise distillation over $M$ beams, BEAM loss forced-scores only two student sequences and reduces the ranking operations from $O(ML^*)$ to $O(L^*)$.
At inference time, the two auxiliary objectives are removed and introduce no additional cost.

\FloatBarrier
 \section{Additional Distillation Analysis}
\label{app:additional-distillation-analysis}

\subsection{Shared-Prefix Distribution}
\label{app:shared-prefix-distribution}

To characterize the availability of target-compatible teacher supervision, Table~\ref{tab:shared-prefix-distribution} reports $P(L^*\geq m)$, where $L^*$ is the longest prefix shared by the target SID and any cached teacher beam.
This cumulative statistic measures the proportion of training examples that retain at least $m$ target-compatible SID levels.

\begin{table}[!htbp]
\centering
\small
\setlength{\tabcolsep}{1.4pt}
\begin{tabular*}{\columnwidth}{@{\extracolsep{\fill}}ccccc@{}}
\toprule
\textbf{Dataset}
& \multicolumn{4}{c}{\textbf{Shared-prefix threshold $m$}} \\
\cmidrule(lr){2-5}
& \textbf{1} & \textbf{2} & \textbf{3} & \textbf{4} \\
\midrule
Beauty & 60.15 & 38.48 & 33.74 & 32.71 \\
Toys   & 69.49 & 40.73 & 35.84 & 34.77 \\
Ad     & 15.88 & 6.14 & 4.87 & -- \\
Video  & 7.58 & 3.01 & 2.75 & -- \\
\bottomrule
\end{tabular*}
\caption{Cumulative shared-prefix retention in the final teacher caches (\%). Column $m$ reports $P(L^*\geq m)$; dashes denote that Kuaishou has no fourth SID level.}
\label{tab:shared-prefix-distribution}
\end{table}
 
Under the selected \texttt{min\_lcp} thresholds of two for Amazon and three for Kuaishou, approximately 40\% of Amazon examples and fewer than 5\% of Kuaishou examples are eligible for prefix-based distillation.
The lower Kuaishou coverage indicates that sufficiently target-compatible teacher beams occur less frequently, making reliability-based filtering particularly important in this domain.
Examples that do not satisfy the threshold continue to receive ground-truth supervision, while the corresponding distillation terms are omitted to avoid introducing weakly aligned teacher signals.

Despite the lower coverage on Kuaishou, the component ablations in Section~\ref{app:kuaishou-additional-ablation} show that restricting supervision to target-compatible teacher prefixes substantially improves performance.
Together, these results indicate that the effectiveness of prefix-based distillation depends more on the reliability of the selected teacher signals than on the proportion of training examples covered by distillation.

\subsection{Learned Hierarchy Mapping}
\label{app:learned-hierarchy-mapping}

Table~\ref{tab:learned-hierarchy-mapping} reports the learned shape parameter $\theta$ and the resulting level scores $f_\theta(d_\ell)$ before normalization over each example's valid shared prefix.
Because Beauty and Toys are jointly trained, they share the Amazon mapping; Ad and Video analogously share the Kuaishou mapping.

\begin{table}[!htbp]
\centering
\small
\setlength{\tabcolsep}{1.4pt}
\begin{tabular*}{\columnwidth}{@{\extracolsep{\fill}}lrrrrr@{}}
\toprule
\textbf{Joint model}
& \textbf{$\theta$}
& \textbf{$f_\theta(d_1)$}
& \textbf{$f_\theta(d_2)$}
& \textbf{$f_\theta(d_3)$}
& \textbf{$f_\theta(d_4)$} \\
\midrule
Amazon ($L=4$)   & 0.998 & 0.321 & 0.606 & 0.834 & 1.000 \\
Kuaishou ($L=3$) & 0.992 & 0.421 & 0.764 & 1.000 & -- \\
\bottomrule
\end{tabular*}
\caption{Learned hierarchy mappings for the joint Amazon and Kuaishou models; the score at the last valid level equals 1 by construction, and the dash denotes that Kuaishou has no fourth SID level.}
\label{tab:learned-hierarchy-mapping}
\end{table}
 
For both joint models, the learned mappings produce distinct level scores that progressively increase with SID depth rather than collapsing to uniform values.
Although the score at the final valid level equals one by construction, the lower scores assigned to earlier levels show that the trained mappings preserve a meaningful separation across the SID hierarchy.
Because exponentiation and prefix-wise normalization preserve this ordering, the resulting SID-loss weights assign progressively greater importance to deeper valid levels.

The different numbers and positions of the intermediate scores arise from the normalized depth $d_\ell=\ell/L$, allowing the same parameterization to accommodate the four-level Amazon hierarchy and the three-level Kuaishou hierarchy.
These learned mappings confirm that \method realizes its intended coarse-to-fine weighting behavior, consistent with the larger teacher advantage observed at deeper SID levels in the main paper.

\FloatBarrier
 
\end{document}